\pdfoutput=1
\documentclass[final,5p,11pt,hidelinks,times]{elsarticle}
\usepackage[final]{microtype}
\usepackage{setspace}

\makeatletter
\let\@afterindenttrue\@afterindentfalse
\makeatother

\usepackage{titlesec}
\titleformat*{\section}{\Large\normalshape\bf}
\titleformat*{\subsection}{\large\normalshape\bf}
\titleformat*{\subsubsection}{\normalshape\bf}

\makeatletter
\renewcommand\paragraph{\@startsection{paragraph}{4}{\z@}%
  {-3.25ex\@plus -1ex \@minus -.2ex}%
  {1.5ex \@plus .2ex}%
  {\normalfont\normalsize\bfseries\itshape}}
\makeatother
\journal{TBD}

\usepackage{hyperref}

\usepackage[font=normalsize,labelfont=bf]{caption}

\usepackage[utf8]{inputenc}

\usepackage{amsmath}
\usepackage{amsfonts}
\usepackage{bm}
\usepackage{siunitx}

\usepackage{tabularx}
\usepackage{booktabs}

\usepackage{xcolor}

\usepackage{graphicx}
\usepackage{subcaption}

\usepackage[status=final,index,multiuser,author=]{fixme}
\fxsetface{margin}{\linespread{1}\scriptsize}
\fxusetheme{color}

\usepackage[displaymath]{lineno}
\newcommand*\patchAmsMathEnvironmentForLineno[1]{%
  \expandafter\let\csname old#1\expandafter\endcsname\csname #1\endcsname
  \expandafter\let\csname oldend#1\expandafter\endcsname\csname end#1\endcsname
  \renewenvironment{#1}%
     {\linenomath\csname old#1\endcsname}%
     {\csname oldend#1\endcsname\endlinenomath}}%
\newcommand*\patchBothAmsMathEnvironmentsForLineno[1]{%
  \patchAmsMathEnvironmentForLineno{#1}%
  \patchAmsMathEnvironmentForLineno{#1*}}%
\AtBeginDocument{%
\patchBothAmsMathEnvironmentsForLineno{equation}%
\patchBothAmsMathEnvironmentsForLineno{align}%
\patchBothAmsMathEnvironmentsForLineno{flalign}%
\patchBothAmsMathEnvironmentsForLineno{alignat}%
\patchBothAmsMathEnvironmentsForLineno{gather}%
\patchBothAmsMathEnvironmentsForLineno{multline}%
}

\begin{document}

\begin{frontmatter}

  \title{
    Hexahedral mesh of anatomical atlas for construction of computational human brain models: Applications to modeling biomechanics and bioelectric field propagation}

\address[UWA]{
  Intelligent Systems for Medicine Laboratory,
  The University of Western Australia,\\
  35 Stirling Highway,
  Perth, WA, Australia}

\address[CRL]{
  Surgical Planning Laboratory,
  Brigham and Women's Hospital,
  Boston, MA, USA}

\address[Harvard]{
  Harvard Medical School,
  Boston, MA, USA}

\author[UWA]{A.T. Huynh\corref{mycorrespondingauthor}}
\ead{andy.huynh@research.uwa.edu.au}
\cortext[mycorrespondingauthor]{Corresponding author}

\author[UWA]{B.F. Zwick}

\author[UWA]{M. Jamshidian}

\author[CRL,Harvard]{M. Halle}

\author[UWA]{A. Wittek}

\author[UWA,Harvard]{K. Miller}

\begin{abstract}

Numerical simulations rely on constructing accurate and detailed models to produce reliable results - a task that is often challenging. This task becomes notably more difficult when the model is of the human brain. We create an anatomically comprehensive hexahedral mesh of the human brain using an open-source digital brain atlas. Digital atlases are valuable tools currently used by medical professionals, medical students, and researchers for gathering, presenting, and discovering knowledge about the human brain. We demonstrate that the atlas can be used to efficiently create an accurate and detailed hexahedral finite element mesh of the brain for scientific computing. We present two case studies. The first case study constructs a biomechanical model of the brain to compute brain deformations and predict traumatic brain injury risk due to violent impact. In the second case study, we construct a bioelectrical model of the brain to solve the electroencephalography (EEG) forward problem, a frequent simulation process used in electrophysiology to study electromagnetic fields generated by the nervous system. We demonstrate efficient and accurate model construction using the meshed anatomical brain atlas, as well as emphasize the importance of effective communication and contextual analysis of results for enabling multi-disciplinary scientific computing research.

\end{abstract}

\begin{keyword}
hexahedral mesh\sep
computer simulation\sep
human brain\sep
digital atlas\sep
biomechanics\sep
bioelectricity

\end{keyword}

\end{frontmatter}

\section{Introduction}
Numerical simulations of the human brain are widely used in medical, scientific, and engineering applications to investigate the brain's complex behavior and responses to various conditions. Developing computational models of the human brain is a difficult task as it has numerous structures and a highly complex geometry. However, technological advancements have allowed for the acquisition, processing, and interconnection of larger amounts of data, making the development of these models increasingly comprehensive and realistic. Interconnecting data from multi-disciplinary fields has expanded the range of practical applications of these models. Some of these models are being developed to assist in medical procedures and practice, neuroscience research and discovery, and brain injury analysis and prevention. Typically, a computational brain model is constructed for a specific application, as each use case has unique requirements and modeling approaches. Although most models use the finite element (FE) method for solving the underlying mathematical equations, they can vary in terms of anatomical accuracy, element types and sizes, and modeling parameters. As a result, the development of FE brain models remains an active area of research, driven by the increasing availability of data opening up new applications and a growing need to create specialized models tailored to each specific use case. 

FE brain models find extensive application in simulating head impact that may cause traumatic brain injury (TBI), one of the leading causes of death and disability worldwide. To reduce the risk of head injury, these models are employed to understand injury mechanisms, determine tolerance thresholds, and develop preventative measures. In a comprehensive review of FE brain models for TBI, Madhukar and Ostoja-Starzewski \cite{madhukar_finite_2019} highlighted 15 prominent models developed between 2003 and 2018. Although these models vary in model description, most discretized the scalp, skull, cerebrospinal fluid (CSF), tentorium cerebelli, falx cerebri, brain tissue and brain stem. The element resolution ranged from approximately 17,000 elements to high-resolution models with around 2.5 million elements. Hexahedral elements were used for 11 of these models, whereas only 4 used tetrahedral elements. It was recommended in this review that the resolution of these FE models should be increased to include finer anatomical features and provide more accurate delineation of material boundaries through direct conversion of image voxels to elements. One notable example that does this is the atlas-based brain model (ABM) derived from the International Consortium of Brain Mapping (ICBM) brain atlas \cite{miller_development_2016}. In this model, each 1 mm isotropic voxel was converted into an element (approx. 2 million elements in total) and included six distinct structures. The cerebrum (combined white and grey matter), cerebellum, CSF and ventricles were extracted from the atlas, while the falx cerebri and tentorium cerebelli were added manually. Although using a brain atlas provides an anatomically accurate mesh for constructing FE brain models, there were noticeable limitations of this study. For example, the model had a staircase (jagged) interface between structures which may cause difficulty for contact and boundary definitions. Meshes generated this way without smoothing the staircase interface were found to have a higher Von Mises stress at the surfaces than in the smoothed models \cite{camacho_improved_1997}. The model also used a high number of elements given that it only extracted a limited number of anatomical structures from the atlas. This results in requiring high computational power for numerical simulations and limits its potential use in time sensitive tasks. We ensured that our mesh had smoothed boundaries where contact or boundary definitions were applied. Additionally, we provide the option of an upsampled mesh, with larger element size and less element count for computationally expensive tasks.

FE brain models are also used in electroencephalography (EEG) and magnetoencephalography (MEG) to study neural activity in the brain. Accurate localization of electrical sources in the brain from EEG and MEG data requires solving the EEG/MEG forward problem, which involves determining surface potentials or external magnetic fields generated by intracranial current sources \cite{schimpf_dipole_2002}. The finite element (FE) method is preferred for solving this problem due to its ability to handle complex volume geometries and model inhomogeneous and anisotropic tissue conductivity \cite{wolters_geometry-adapted_2007}. Modern imaging technologies have enabled the use of detailed, realistic head models, replacing simplified spherical models. Efficient construction of accurate FE head models is crucial in neuroscience research and clinical applications, such as studying brain function \cite{lopesdasilva_eeg_2013} and localizing epileptic seizures \cite{barkley_meg_2003}. There are widely available open-source software for automatic generation of realistic head models from individual MRI scans \cite{tran_improving_2020, makarov_simnibs_2019, huang_realistic_2019, oostenveld_fieldtrip_2011}. Some of these software incorporate digital brain atlases, but are limited in the number of anatomical structures. Additionally, most software generates either tetrahedral meshes or voxel-based hexahedral meshes with staircase interfaces. Studies have shown that voxel-based hexahedral meshes with node-shifted surface smoothing can reduce EEG forward modeling errors compared to tetrahedral meshes \cite{wolters_geometry-adapted_2007}.

In this paper, we present a comprehensive approach to hexahedral meshing of a digital human brain atlas, including important features such as conforming boundaries and anatomically labeled elements. Digital atlases are powerful tools commonly used by medical professionals, medical students, and researchers for gathering, presenting, and discovering knowledge about the human body. They are continuously evolving in terms of content, applications, functionality and availability by many big and well-funded projects \cite{nowinski_evolution_2021}. Our study utilizes a digital brain atlas from the Open Anatomy Project (OAP) \cite{halle_open_2017}, an open-source project offering several advantages for efficient construction of accurate FE models. The OAP hosts a diverse library of detailed digital anatomical atlases, including brain, thorax, liver, inner ear, head and neck, knee, and abdominal atlases. This library allows our approach to be applied across various anatomical regions. Additionally, the OAP's collaborative development model, involving anatomists, researchers, illustrators, and teachers around the world, enables customization and expansion of its already extensive library of atlases \cite{open_anatomy_project_open_2024}. We aim to create labeled, finite element hexahedral meshes of these anatomical atlases, starting with the brain atlas, and contribute them to the OAP platform, thereby advancing multi-disciplinary scientific computing research.

Our approach converts the OAP's SPL/NAC brain atlas to an anatomically accurate FE mesh using primarily open-source software. For the atlas to be useful for constructing FE models, we add missing structures including the skull, scalp, and cerebrospinal fluid (CSF), while merging existing structures to form grey and white matter. We developed SlicerAtlasEditor \cite{huynh_sliceratlaseditor_2023}, an extension for 3D Slicer \cite{fedorov_3d_2012}, for this purpose. This extension uses the Human Atlasing Working Group (HAWG) data structure, allowing seamless authoring of OAP's digital anatomical atlases tailored to specific needs. We extract triangulated surfaces for interfaces requiring smooth boundaries, such as the scalp-skull and skull-CSF interfaces. A conforming hexahedral mesh is generated using Coreform Cubit's \cite{coreform_llc_coreform_2018} overlay-grid method. This method allows specification of the base Cartesian mesh size with which selected nodes near boundaries are shifted to conform to triangulated surfaces, then smoothed to improve element quality \cite{owen_sculpt_2019}. We can specify the underlying atlas voxel size to be directly converted to same-size elements, minimizing interpolation errors. A coarser base mesh option is available for faster computation, though at the cost of reduced anatomical and numerical accuracy. After mesh generation, we label elements with corresponding anatomical labels using MVox \cite{zwick_mvox_2020} and the Visualization Toolkit (VTK) \cite{schroeder_visualization_2006}. The meshed atlas is then converted to formats compatible with MFEM \cite{anderson_mfem_2021} and Ansys LS-DYNA \cite{ansys_ansys_2023}. We present two case studies in computational modeling: brain injury mechanics, where we simulate brain impact using linear and angular acceleration time histories acquired from human head cadaver studies \cite{hardy_study_2007, hardy_investigation_2001}; and bioelectric field propagation, demonstrating the meshed atlas's versatility by solving the EEG forward problem using a simple dipole model, as applied in Zwick et al. \cite{zwick_patient-specific_2022}. In summary, we emphasize the importance of efficient and accurate model construction using the meshed anatomical brain atlas, as well as effective communication and contextual analysis of results linked to anatomical structures. This approach integrates the meshed atlas with computational modeling and simulation, enabling a seamless workflow for model creation, communication, analysis, and visualization in a relevant disciplinary context.
\section{Hexahedral meshing of anatomical atlas}
\subsection{Digital anatomical atlas}
\begin{figure*}
\centering 
\includegraphics[width=\textwidth]{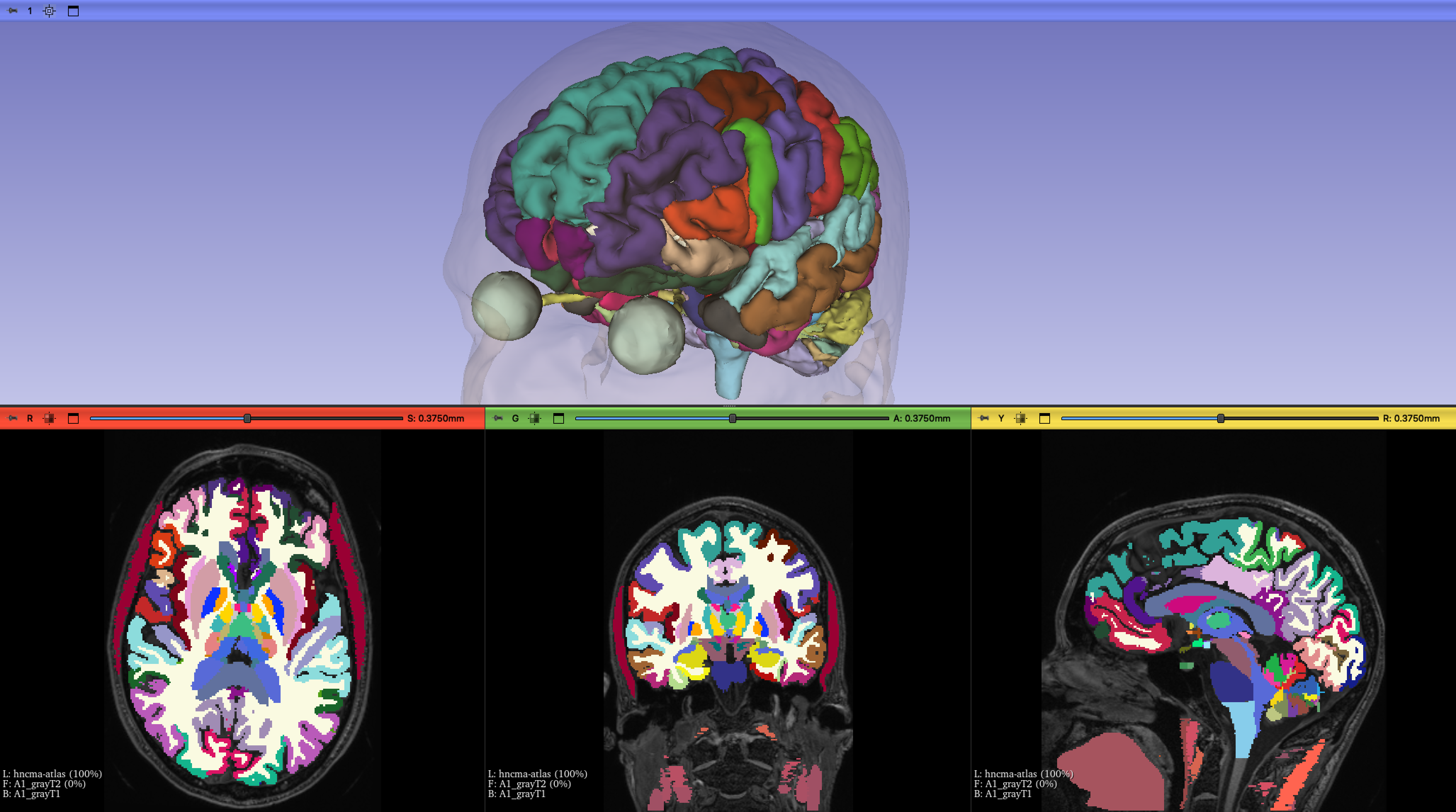} 
\caption{Open Anatomy Project's SPL/NAC Brain Atlas. Top is the 3D rendering of the label map and the bottom is the label map overlaid onto the MRI scan} 
\label{fig1} 
\end{figure*}

The SPL/NAC digital brain atlas, part of the Open Anatomy Project, is a collaborative effort between Brigham and Women's Hospital and Massachusetts General Hospital \cite{halle_m_multi-modality_2017}. This MRI-derived atlas is based on a healthy 42-year-old male volunteer and comprises more than 300 labelled anatomical structures, representing over two decades of development. Data acquisition took place at the Martinos Center for Biomedical Imaging using a Siemens 3T scanner with a multi-array head coil. The dataset includes volumetric whole head MPRAGE (T1-weighted) and T2-weighted MRI series, both with 0.75 mm isotropic voxel size. The atlas features per-voxel labeling of anatomical structures based on downsampled 1 mm isotropic resolution volumes, with a structural hierarchy loosely based on the Radlex ontology. Label generation involved Freesurfer's automatic parcellation \cite{fischl_freesurfer_2012} followed by extensive manual segmentation. For our finite element mesh creation, we utilize the 1 mm isotropic resolution volumetric whole head MRI scans and corresponding anatomical labels (label map). Figure \ref{fig1} shows the 3D rendering of the label map and the volumetric view of the label map overlaid with the original MRI scan. This comprehensive atlas, along with other anatomical atlases, is available for viewing and free download on the OAP platform \cite{open_anatomy_project_open_2024, halle_open_2017}.

\subsection{Pre-processing}
The OAP's SPL/NAC brain atlas currently lacks key structures necessary for constructing comprehensive FE models for our case studies. To address this limitation, we developed a process to generate an additional 'Material label map' derived from the original atlas. The 'Material label map' is a 3D volume that contains discrete "label" values corresponding to anatomical structures essential for defining boundary conditions, material properties, and contact interfaces in FE modeling. To create this label map, we developed SlicerAtlasEditor \cite{huynh_sliceratlaseditor_2023}, an extension for the open-source medical research software 3D Slicer \cite{fedorov_3d_2012, pieper_na-mic_2006}. Our extension allows users to modify any of OAP's anatomical atlases by applying boolean operations to add, merge or remove labels. We used this, along with open-source skull-stripping tools, to incorporate crucial structures including the scalp, skull, cerebrospinal fluid (CSF), grey matter, white matter, and the ventricular system.

\subsubsection{Scalp, skull and cerebrospinal fluid labels}
To create the scalp, skull, and CSF labels, we used SynthStrip, a deep learning skull-stripping tool \cite{hoopes_synthstrip_2022}. The tool incorporates a model trained using a deep convolutional neural network on various images synthesized with a deliberately unrealistic range of anatomies, acquisition parameters, and artifacts. The resulting model is able to generalize across imaging modalities, anatomical variability, and acquisition schemes. We decided to use this tool as it consistently outperformed classical skull-stripping software when measured against quantitative metrics such as mean DICE score, mean surface distance, Hausdorff distance, and volume distance. We applied this tool to extract a binary brain mask from the atlas' MRI scan, which captures both the brain tissue and surrounding CSF. For our purpose, the scalp and skull labels were created using a simplified process, by offsetting the skull-stripped brain mask by 4 mm (4 voxels).

\subsubsection{Grey matter, white matter and ventricular system labels}
We used the SlicerAtlasEditor extension \cite{huynh_sliceratlaseditor_2023} to create grey matter, white matter and ventricular system labels, derived from the OAP's SPL/NAC brain atlas \cite{huynh_sliceratlaseditor_2023}. The OAP's SPL/NAC brain atlas follows the Human Atlasing Working Group (HAWG) data format \cite{human_atlasing_working_group_standardizing_2016}, which is a standardized format of metadata describing the atlas structure. The HAWG format is written as a JSON file, capturing information about the relationships between different anatomical structures and groups \cite{halle_open_2017}. Our atlas editing extension performs Boolean operations to remove, group, and merge different anatomical labels using this HAWG format. To facilitate material property assignment, we grouped and merged all the anatomical structures comprised of grey matter, white matter and the ventricular system. Additionally, we removed the skin and muscle labels from the label map as these were not used in the simulation.

\subsubsection{Material label map}
\begin{figure*}
\centering 
\includegraphics[width=\textwidth]{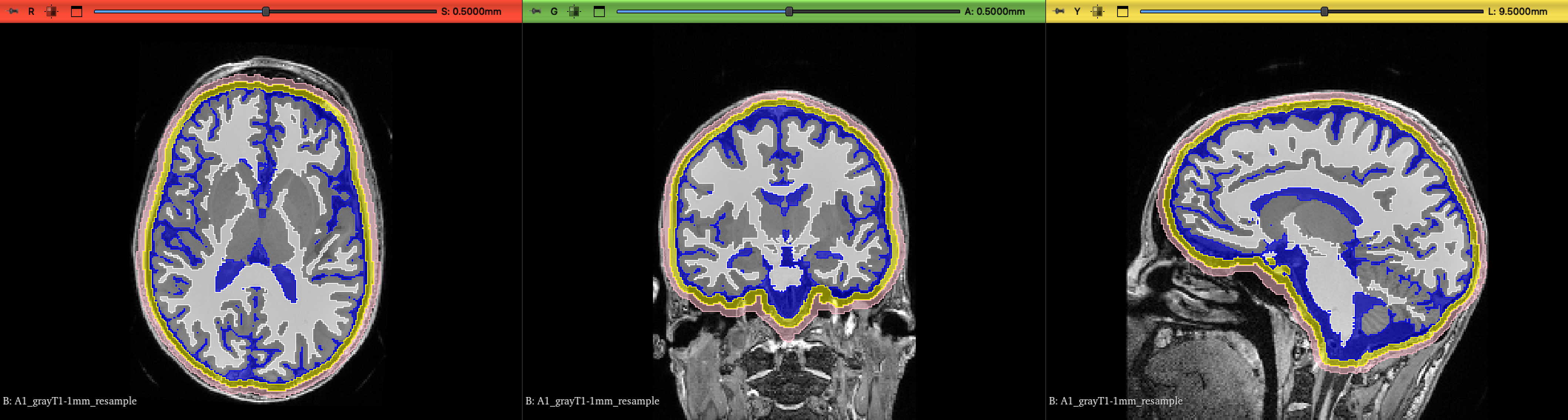} 
\caption{The Material label map consisting of the scalp, skull, CSF, grey matter and white matter derived from the SPL/NAC brain atlas using the SlicerAtlasEditor extension. The label map is color coded with pink as the scalp, yellow as the skull, blue as the CSF and ventricles, grey as the grey matter and white as the white matter.} 
\label{fig2} 
\end{figure*}

The labels created above were merged together into a single label map called the 'Material label map'. The material label map consists of the scalp, skull, CSF, grey matter, white matter and the ventricular system. This is presented in Figure \ref{fig2}. Our process simplifies the customization of OAP's anatomical atlases to suit user's FE modeling requirements.

\subsection{Mesh generation}
\subsubsection{Surface extraction}
Surface meshes are used as boundaries to generate grid-based hexahedral mesh with smooth interfaces. The surfaces were generated using a marching cubes algorithm \cite{lorensen_marching_1987} applied on the scalp, skull and CSF labels. This ensures we have smooth boundaries for applying boundary conditions and defining surface contact when constructing the FE model. We created surface meshes for the outer scalp, scalp-skull interface and skull-CSF interface.

\subsubsection{Conforming, grid-based hexahedral mesh generation}
To create a conforming hexahedral volumetric mesh, we used the Sculpt application \cite{owen_sculpt_2019} provided in the commercial software, Coreform Cubit \cite{coreform_llc_coreform_2018}. The application generates conforming hexahedral meshes automatically using the overlay-grid (or mesh-first) procedure. The procedure begins with a Cartesian grid as the base mesh, with surface mesh/meshes imposed to capture the geometric description. The nodes from the base grid that are near the boundary are projected to the boundary of the surface which distorts the hexahedral elements. A pillow layer is inserted at the surface and smoothing procedures are employed to improve the mesh quality of these elements that are around the boundary region.

\begin{figure*}
\centering 
\includegraphics[width=\textwidth]{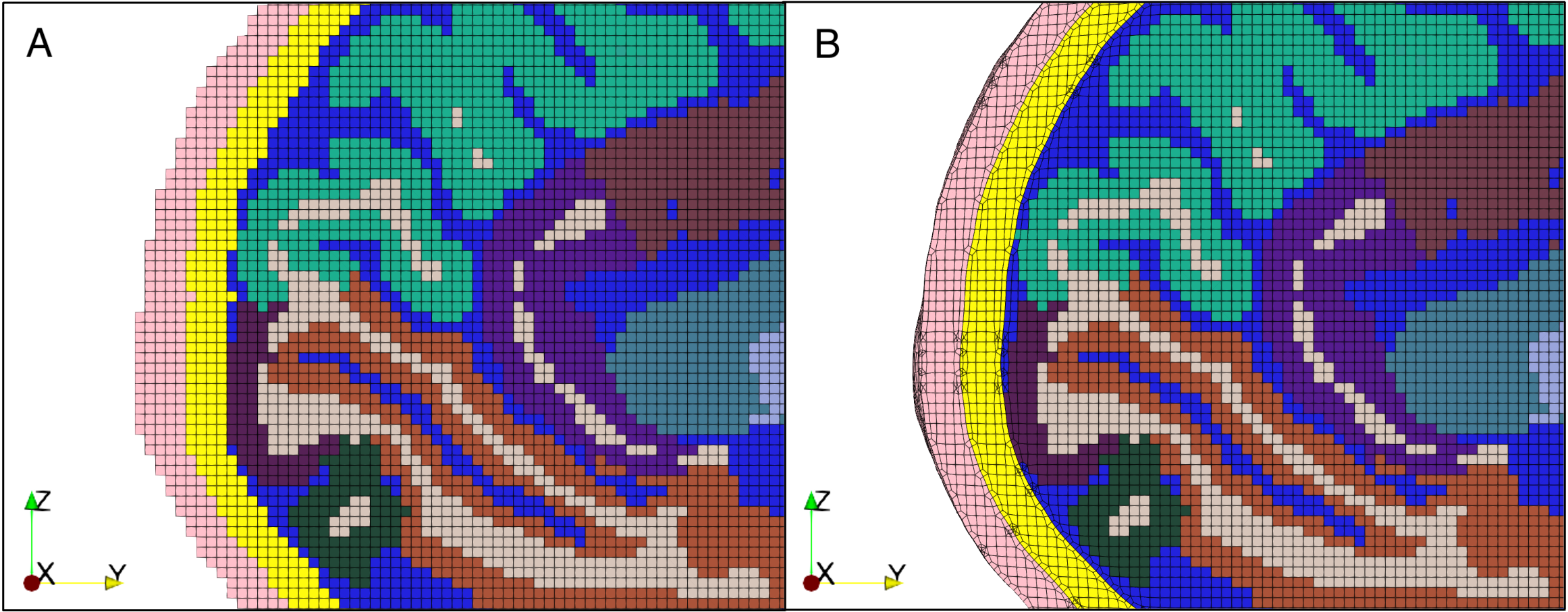} 
\caption{Close-up sagittal plane cut of the labelled meshed anatomical brain atlas showing smoothing of boundaries by the Sculpt meshing procedure. (A) Close-up of the mesh without smooth boundaries and (B) close-up of the mesh with smooth boundaries at the outer scalp, scalp-skull and skull-CSF interface.} 
\label{fig3} 
\end{figure*}

We defined the Cartesian grid by specifying a bounding box with regular intervals. For our biomechanical model, we used a bounding box with the same size $256 \times 256 \times 256$ ($1 \times 1 \times 1$ mm) as the isotropic MRI image and label map volumes used for OAP's SPL/NAC brain atlas. For our bioelectric model, we used a bounding box with size $128 \times 128 \times 128$ ($2 \times 2 \times 2$ mm), which is the up-sampled size of the brain atlas. The overlay-grid procedure used in our method allows for user-defined element sizes. We show that our mesh can be created using the original size of the brain atlas voxels or by a reduced size for computational efficiency. The triangulated surface meshes extracted from the scalp, skull and CSF labels were used for the meshing procedure to conform to the geometric boundaries. Figure \ref{fig3} shows a close-up of the labelled meshed anatomical brain atlas before and after using the overlay grid procedure with boundary smoothing \cite{owen_sculpt_2019, coreform_llc_coreform_2018}.

\subsection{Mesh quality}
We checked the mesh quality for two meshes with 1:1 and 8:1 voxel-to-element ratios, used for the biomechanical and bioelectric model respectively. The reported metrics included the scaled jacobian, aspect ratio and skew metric. These metrics were computed using VTK library's implementation of the Verdict library \cite{stimpson_verdict_2007}. The scaled jacobian is defined as the minimum determinant of the jacobian matrix evaluated at each corner and the center of the element, divided by the corresponding edge lengths. The aspect ratio is defined as the ratio of the longest to shortest edge. The skew metric computes the degree to which a pair of vectors are parallel using the dot product, and takes the maximum.

\begin{figure*}
\centering 
\includegraphics[width=\textwidth]{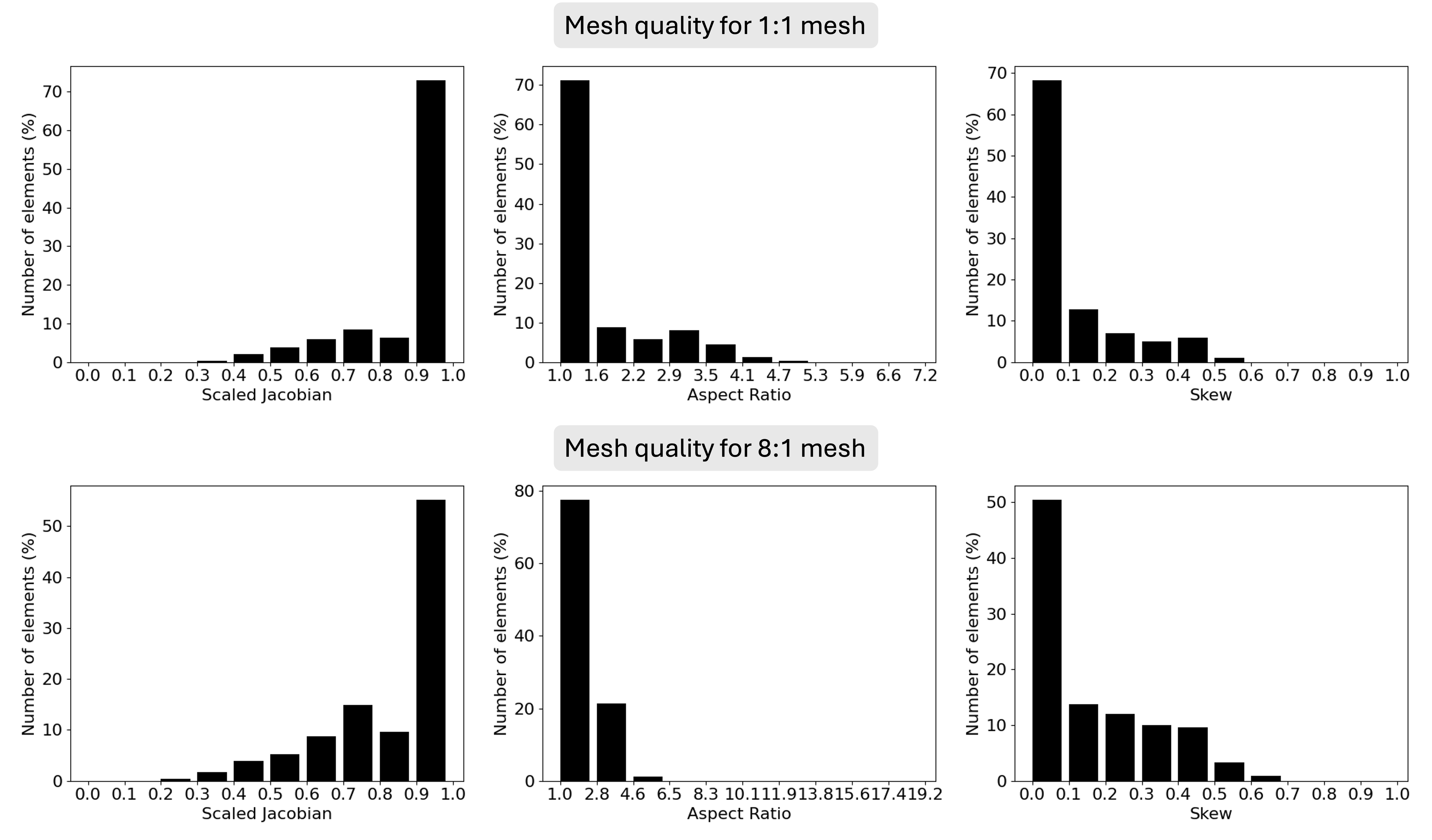}
\caption{Distribution of mesh quality metrics for 1:1 and 8:1 voxel-to-element ratio meshes. The top row presents mesh quality for the 1:1 ratio mesh, while the bottom row shows results for the 8:1 ratio mesh. For each mesh, the horizontal axis represents the mesh quality metric, and the vertical axis displays the percentage of elements within each quality range. } 
\label{fig4} 
\end{figure*}

Both the 1:1 and 8:1 meshes exhibit high quality for the majority of the elements. It should be noted that there are no universally accepted, minimally required quality standards for an FE mesh as there are multiple ways to judge element quality \cite{yang_isoparametric_2018}. There are, however, general recommendations by experienced engineers. Both recommendations from the Verdict library \cite{stimpson_verdict_2007} and King Yang \cite{yang_isoparametric_2018}, were used as guidance for generating a high-quality hexahedral FE mesh. For the 1:1 mesh, the percentage of elements with scaled jacobian $>$ 0.5 was $97.5\%$; the percentage of elements with aspect ratio $<$ 3 was $88.0\%$; and the percentage of elements with skew $<$ 0.5 was $99.0\%$. Scaled jacobian ranged from 0.210 to 1 with an average of 0.910, aspect ratio ranged from 1 to 7.17 with an average of 1.57 and skew ranged from 0 to 0.671 with an average of 0.0950. For the 8:1 mesh, the percentage of elements with scaled jacobian $>$ 0.5 was $97.5\%$; the percentage of elements with aspect ratio $<$ 3 was $88.0\%$; and the percentage of elements with skew $<$ 0.5 was $99.0\%$. Scaled jacobian ranged from 0.213 to 1 with an average of 0.950, aspect ratio ranged from 1 to 19.2 with an average of 1.28 and skew ranged from 0 to 0.652 with an average of 0.00560. The distribution of the mesh quality metrics for both meshes are presented in Figure \ref{fig4}.

\subsection{Mesh labelling}
\begin{figure*}
\centering 
\includegraphics[width=\textwidth]{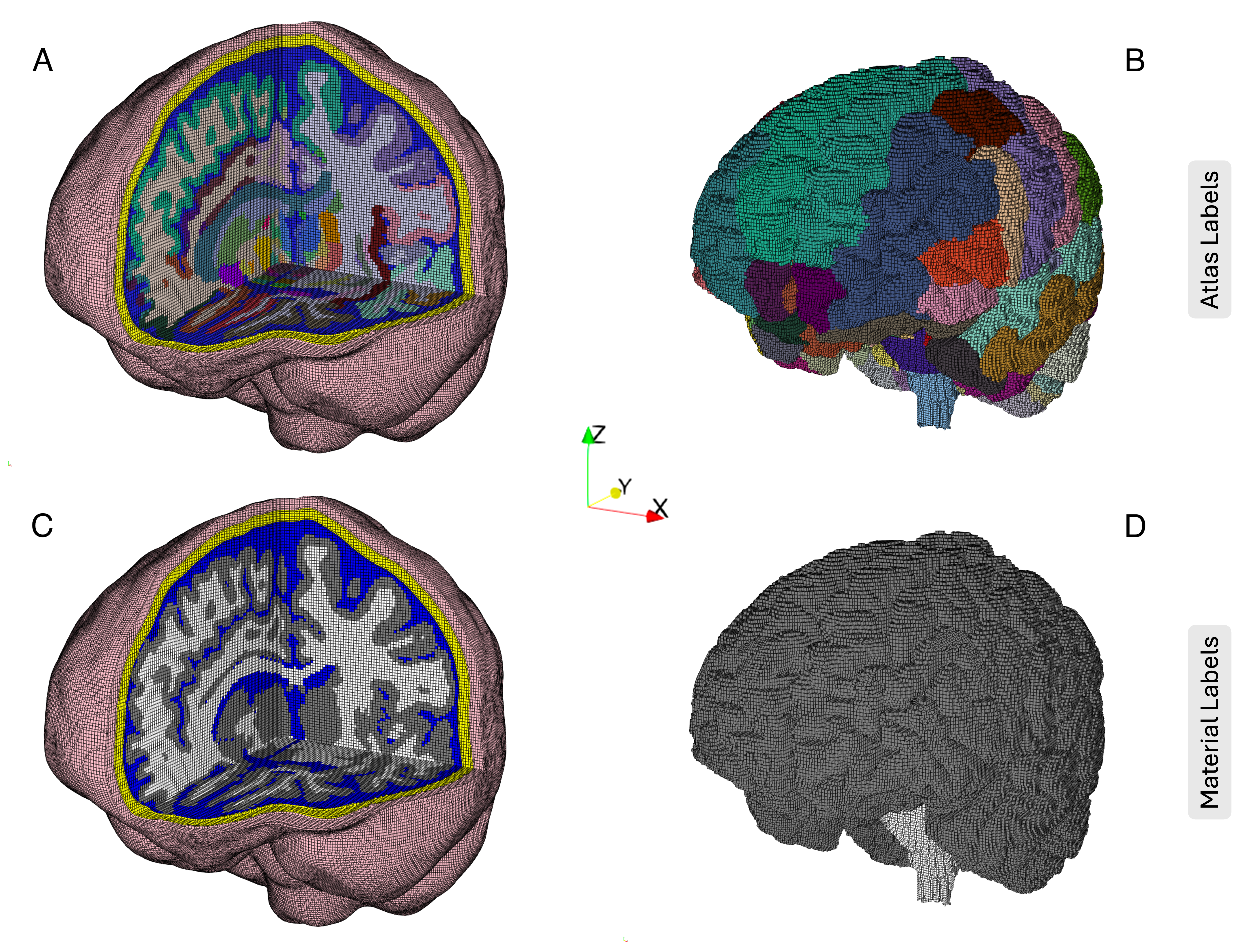} 
\caption{The meshed anatomical brain atlas. Top row: (A) 3D cut-out view of the mesh with OAP's SPL/NAC brain atlas labels and (B) 3D view of the mesh without the skull and scalp. Bottom row: (C) 3D cut-out view of the mesh material labels for constructing finite element (FE) models and (D) 3D view of mesh without the skull and scalp.} 
\label{fig5} 
\end{figure*}

We used MVox \cite{zwick_mvox_2020} and the VTK library \cite{schroeder_visualization_2006} to label the generated grid-based hexahedral mesh with the anatomical and material labels from the OAP's SPL/NAC brain atlas. First, MVox was used to create a VTK unstructured grid from the original atlas label map and material label map. The grid-based hexahedral mesh generated using Coreform Cubit was also converted to a VTK unstructured grid. This is to ensure both datasets had compatible data format when interacting with functions in the VTK library. To efficiently match cells between our label maps and the hexahedral mesh, we employed a VTK function called vtkCellLocator.FindClosestPoint. This function utilizes an octree-based spatial search method, a technique for quickly locating cells in three-dimensional space. The method works by dividing the 3D space into a hierarchical structure of cubic regions, or octants. Each octant is marked as either empty or containing cells, with the smallest divisions (leaf octants) storing lists of the cells they contain. This octree structure enables rapid searching and assigning of labels to the mesh. This function is especially useful when adding additional anatomical structure labels from the digital atlas, or addition of a different digital atlas such as a functional atlas for visualisation and analysis of computed results. The labelled meshed anatomical brain atlas is presented in Figure \ref{fig5}, showing both the anatomical and material labels (Note: both are the same mesh consisting of both labels).

\clearpage
\section{Case study I: Biomechanics}
Finite element (FE) brain models are essential tools in biomechanics for investigating the mechanical response of the brain during impact, playing a crucial role in traumatic brain injury research. Many studies in this field rely on established FE brain models developed by different groups, characterized by predefined numerical settings, specific geometries, limited anatomical structures, and software-specific implementations \cite{giudice_analytical_2019}. While valuable, these models are often difficult to adjust and customize due to their complexity. To address these limitations, we present a novel workflow for constructing biomechanical models using our meshed anatomical atlas. This approach offers several advantages over existing models. Our meshed anatomical atlas simplifies brain model construction by providing users with meshes of varying size (or density), enabling researchers to balance accuracy and computational efficiency. The mesh is comprehensively labeled with anatomical structures from the OAP's SPL/NAC brain atlas, facilitating efficient assignment of material properties, boundary conditions, and surface contact definitions. Furthermore, they enable advanced queries of computed results and their statistics for context-based analysis. Our meshed anatomical atlas is compatible with various commercial and open-source software platforms, ensuring broad applicability across research environments. Our approach addresses the limitations of current models, offering a more flexible and customizable solution for brain injury modeling.

\subsection{Model description}
We present a novel workflow for constructing a biomechanical model of the brain using our meshed anatomical atlas. Our approach leverages anatomically labeled elements and smooth boundary interfaces, offering several advantages in model development and analysis. Our biomechanical model aims to predict brain deformation and traumatic brain injury risk due to violent impact. The model construction is based on approaches outlined by Wang et al. \cite{wang_prediction_2018}, which builds upon the well-established and validated THUMS Version 4.0 human body model by Toyota \cite{shigeta_development_2009, watanabe_research_2011} with minor modifications. We replicate the experiments conducted by Hardy et al. \cite{hardy_investigation_2001}, who examined six scenarios using two post-mortem human subjects (three experiments each). These experiments applied various levels of linear and angular acceleration, representing different impact locations (frontal and occipital). We define the loading in our model using the head linear and angular acceleration-time histories around all three axes of the coordinate system associated with the head's center of gravity (COG). To demonstrate the model's flexibility, we present different representations of the ventricular system and compare their results. Additionally, we compare our model's results with Wang et al. \cite{wang_prediction_2018}. We implement the model and simulation using Ansys LS-DYNA explicit dynamics nonlinear finite element code \cite{ansys_ansys_2023}. 

\subsection{Modelling approaches}

\subsubsection{Brain-skull interface}
For the brain-skull interface, we followed recommendations in the study by Wang et al. \cite{wang_prediction_2018}. The study found that the approach used to model the brain skull interface in the THUMS Version 4.0 model, which includes direct representation of the subarachnoidal surface with CSF, resulted in the smallest differences with the experimental results by Hardy et al. \cite{hardy_investigation_2001}. More specifically, the model includes the direct representation of the CSF, dura mater, arachnoid and pia mater which is consistent with the anatomical structure of brain-skull interface according to Haines et al. \cite{haines_subdural_1993}. This approach allows for relative movement between the brain and skull, while preventing separation/detachment. Material properties used for the brain-skull interface used in the THUMS model can be found in Table \ref{tab1}.

\begin{table*}[h!]
\caption{Material properties used in the brain-skull interface in THUMS Version 4.0 model and applied to our model.}
\begin{tabular*}{\textwidth}{@{\extracolsep\fill}lllll}
    \toprule
    Part    &   Mass density ($\text{kg/m}^3$) &   Young's modulus (MPa)    &   Poisson ratio  &   Thickness (mm)  \\
    \midrule
    Cerbral spinal fluid (CSF)   &   1000   &   1.6E-4   &   0.49   &      \\
    Pia mater &   1000   &   1.1   &   0.40   &   0.40   \\
    Arachnoid    &   1000   &   1.1   &   0.40   &   0.40   \\
    Dura mater  &   1133   &   70   &   0.45   &   1   \\
    \bottomrule
\end{tabular*}
\label{tab1}
\end{table*}

\subsubsection{Constitutive model and properties of brain tissue}
In THUMS Version 4.0, the brain paranchyma is modelled using a linear viscoelastic constitutive model, however, numerous studies have indicated the brain tissue exhibits the behaviour best represented using hyperelastic or hyperviscoelectic constitutive models. As such, we use an Ogden hyperviscoelastic model as described by Miller and Chinzei and Mihai et al. \cite{miller_constitutive_1997, miller_mechanical_2002, wittek_subject-specific_2008, mihai_comparison_2015} to define the material properties of grey matter and white matter in both the cerebrum and cerebellum:

\begin{equation}
\label{eq1}
W = \frac{2}{\alpha^2}\int_{0}^{t}\Big[G(t - \tau)\frac{d}{d_\tau}(\lambda_{1}^{\alpha} + \lambda_{2}^{\alpha} + \lambda_{3}^{\alpha} - 3)\Big]d\tau + K(J - 1 - \ln(J))
\end{equation}

\begin{equation}
\label{eq2}
G(t) = G_i + (G_0 - G_i)e^{\frac{-t}{\tau}},
\end{equation} \\
where $W$ is the potential function, $\lambda_i (i = 1, 2, 3)$ are the principal stretches, $G_0$ is an instantaneous shear modulus, $G_i$ is the relaxed shear modulus, $\tau$ is the characteristic time, $\alpha$ is the material coefficient which can assume any real value without restrictions \cite{wittek_patient-specific_2007}, $K$ is the bulk modulus, and $J$ is the relative volume change.

\begin{table*}[h!]
\caption{Ogden hyperviscoeleastic model parameters for brain tissue used in this study}
\begin{tabular*}{\textwidth}{@{\extracolsep\fill}lllll}
    \toprule
    Part    &   $G_0$ (Pa) &   $\alpha$    &   $G_i$ (Pa)  &   $\tau$ (s)  \\
    \midrule
    White matter   &   1100   &   -4.7   &   550   &   0.06   \\
    Grey matter    &   850   &   -4.7   &   425   &   0.06   \\
    \bottomrule
\end{tabular*}
\label{tab2}
\end{table*}

The 'Material label map' we created earlier combined the individual anatomical labels in the OAP's SPL/NAC brain atlas into grey matter and white matter. The label map was used to label the corresponding mesh elements, which are used as a reference for defining the material properties in Table \ref{tab1} and \ref{tab2}. In Figure \ref{fig5} (C) and (D), we highlight the labels from the 'Material label map' of the meshed anatomical atlas used to define the material properties. 

\subsubsection{Constitutive model and properties of ventricles}
\begin{figure*}[h!]
\centering 
\includegraphics[width=0.85\textwidth]{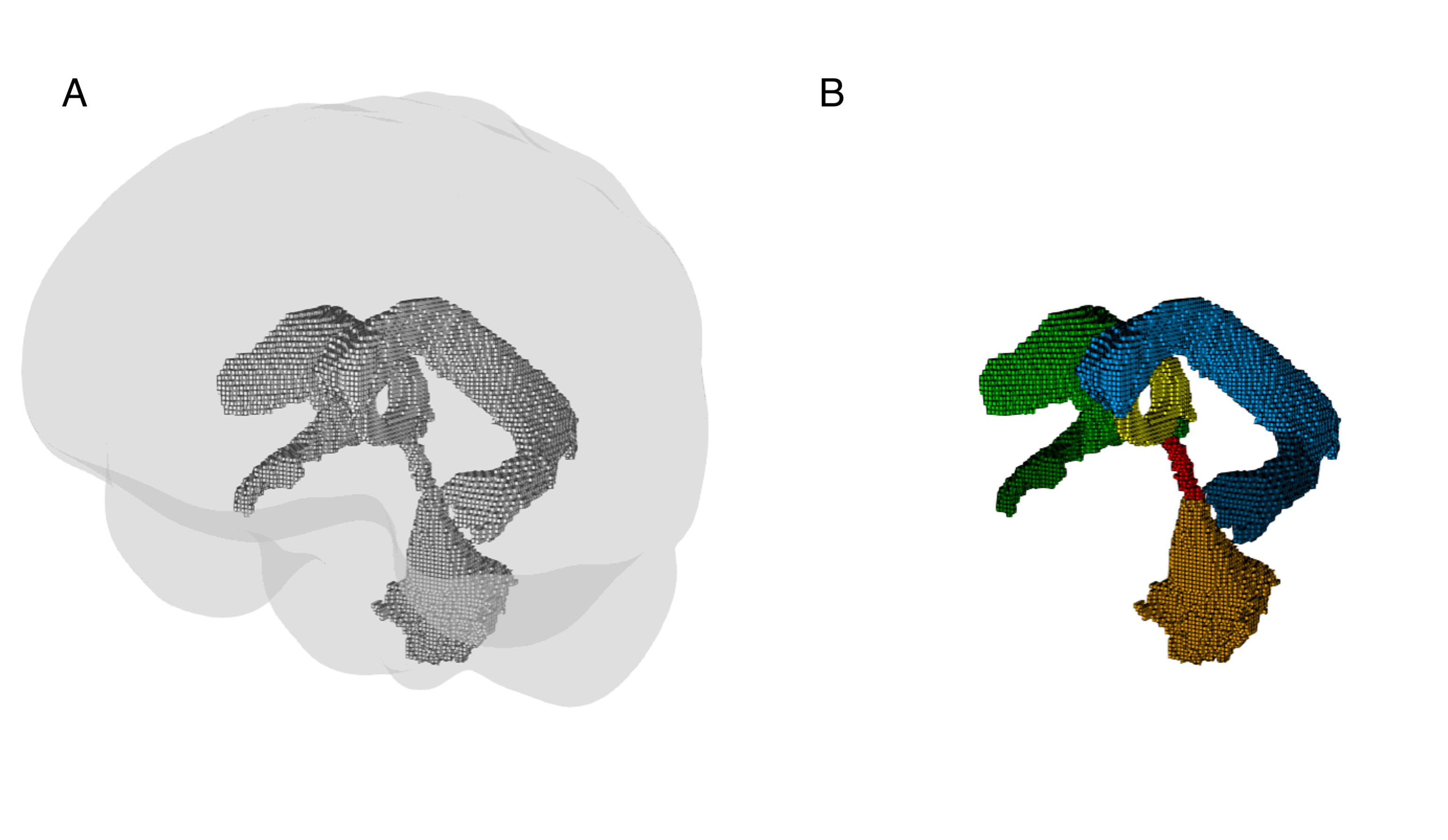}
\caption{(A) 3D model shows the meshed anatomical brain atlas with the ventricular system highlighted. (B) 3D model shows the ventricular system with its substructures, including the left lateral ventricles (light blue), temporal horn of left lateral ventricles (dark blue), right lateral ventricles (light green), temporal horn of right lateral ventricles (dark green), third ventricle (yellow), aqueduct (red) and the fourth ventricle (orange). The labelling of the ventricular system in the FE mesh was used to construct three different models. Model (A): elastic-fluid ventricles; Model (B): homogeneous parachnyma; Model (C): empty cavity ventricles.} 
\label{fig6} 
\end{figure*}

We present different modelling approaches for representing the ventricles. We construct and solve three different models. Model A is the complete model which includes a geometrically accurate representation of the ventricular system, modelled as an elastic-fluid material. Elastic-fluid material is an option that can be set for elastic material to exhibit fluid-like behaviour \cite{ansys_ls-dyna_2021}. The ventricular system is included in the OAP's SPL/NAC brain atlas and as a result, our meshed anatomical brain atlas. It  comprises of the left/right lateral ventricles, temporal horn of left/right lateral ventricles, third ventricle, aqueduct and the fourth ventricle. Model B is modelled as in THUMS Version 4.0 which does not separate the ventricles from the brain paranchyma and instead models it homogenously as part of the white matter tissue. Model C models the ventricles as an empty cavity, which is a simplification used to remove the need for meshing the ventricles. We show that with the meshed anatomical brain atlas, construction of biomechanical brain models with different modelling approaches can be done seamlessly without any additional manual segmentations of the internal structures of the brain, a time-consuming process that was required in our past models \cite{wittek_effects_2011, bilston_computational_2011}. We present the ventricular system used in the FE models in Figure \ref{fig6}.

\subsection{Results of biomechanical case study}
We present results similar to those shown in Wang et al. \cite{wang_prediction_2018}, including quantile plots of the maximum principal strain and shear strain (Almansi) at the time when maximum strain value was observed. Additionally, we show the maximum principal strain (MPS). The MPS is typically used as a brain tissue deformation-based injury criteria to predict the possibility of diffuse axonal injury (DAI). We compare the MPS across three different modeling approaches for ventricles used in literature (Model A, B and C), as well as with results presented in Wang et al. \cite{wang_prediction_2018} (Reference Model). While we show results for the overall brain typically presented in the field of brain injury, our meshed anatomical brain atlas allows for analysis of local brain regions to better understand neuropathological effects. This is achieved by linking computed results with the anatomical structures made available from OAP's SPL/NAC brain atlas, enabling multi-disciplinary analysis. For example, focal brain injury often occurs as a result of traumatic brain injury (TBI), and exploring the relationship between white matter structure and cognitive function following TBI is important, as cognitive deficits are commonly observed after TBI due to diffuse axonal injury \cite{kinnunen_white_2011}. We present results of principal and shear strain at important anatomical regions associated with diffuse axonal injury reported in literature, including the corpus callosum and brain stem components such as the midbrain, pons, and medulla oblongata \cite{johnson_axonal_2013, gentry_imaging_1994}.

\begin{figure*}
\centering 
\includegraphics[width=\textwidth]{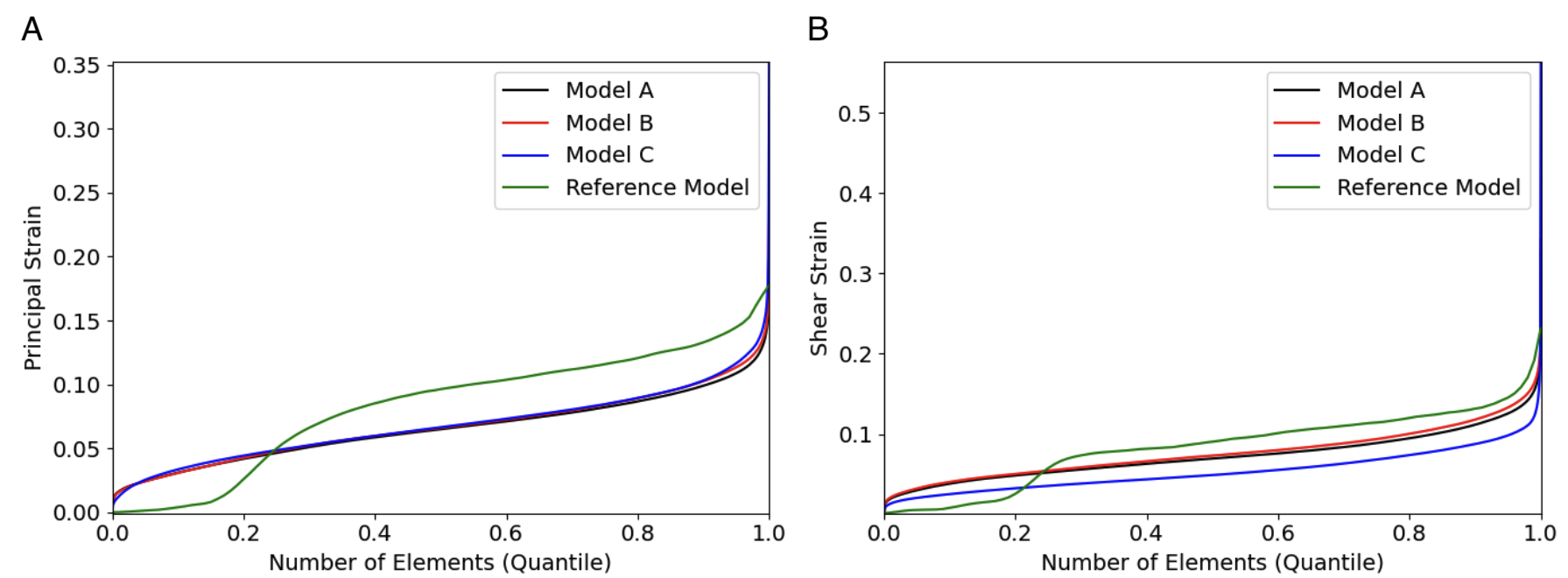}
\caption{Quantile plots of the principal (left) and shear (right) strain (Almansi) of the brain paranchyma at the time where the maximum strain was observed in the simulation. The plots show the observed strains for different modelling approaches for the ventricular system and the results found in Wang et al. \cite{wang_prediction_2018}. Model (A): elastic-fluid ventricles; Model (B): homogeneous paranchyma; Model (C): empty cavity venctricles; Reference Model: THUMS model used in Wang et al.\cite{wang_prediction_2018}} 
\label{fig7} 
\end{figure*}

In Figure \ref{fig7}, quantile plots of the principal and shear strain (Almansi) of the brain paranchyma at the time where the maximum strain was observed. We present results for the different modelling approaches for the ventricular system (Model A, B and C) and the results found in Wang et al. \cite{wang_prediction_2018} (Reference Model). Our results show that there is negligable differences in principal strain (Almansi) between Models A, B and C. For quantiles 0.25, 0.5, 0.75 and 0.95, the respective principal strains for the three models are approximately 0.048, 0.066, 0.084 and 0.113. Shear strain (Almansi) show noticeable differences between Model C and the rest (Model A and B). Comparing the shear strain of Model A and Model B with Model C at the respective qunatiles; shear strains for Model A and B are approximately: 0.053, 0.071, 0.091 and 0.1148, and for Model C: 0.036, 0.049, 0.068 and 0.088. The difference between the shear strains at the respective quantiles are 0.017, 0.022, 0.023 and 0.027. Comparing our results to Wang et al. \cite{wang_prediction_2018} (Reference Model), there are larger differences in both principal strain and shear strain. For both the principal strain and shear strain plots, we observe that the reference model experience smaller strain values at the 0.25 quantile and larger strain values beyond this point. 

\begin{figure*}
\centering 
\includegraphics[width=0.6\textwidth]{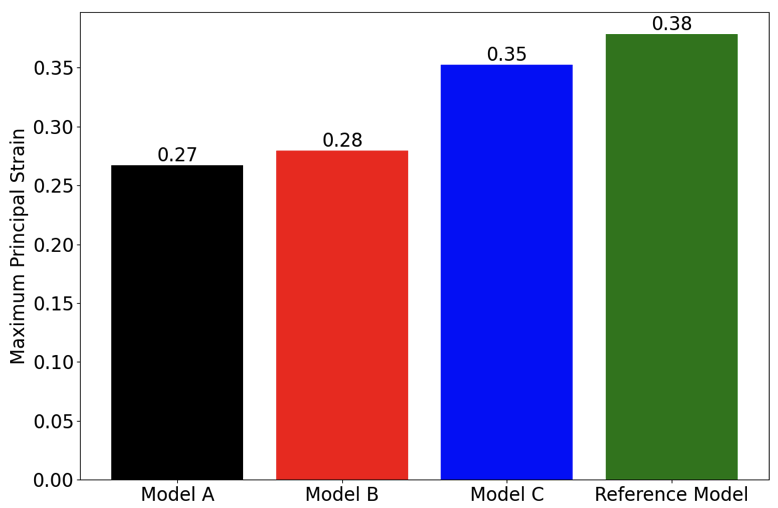}
\caption{Comparison of Maximum principal strain (Almansi). Model (A): elastic-fluid ventricles; Model (B): homogeneous paranchyma; Model (C): empty cavity venctricles; Reference Model: THUMS model used in Wang et al. \cite{wang_prediction_2018}} 
\label{fig8} 
\end{figure*}

Figure \ref{fig8} presents the MPS (Almansi) typically used as brain injury criteria. We compare results from different modeling approaches for the ventricular system (Models A, B, and C) and the results from Wang et al. \cite{wang_prediction_2018} (Reference Model). The maximum principal strain (Almansi) for Model C was greater than both Models A and B, with Models A and B exhibiting negligible differences. Comparing Model C with Models A and B, a difference in MPS of 0.08 and 0.07 was found, respectively, while only a minor difference of 0.01 was found between Models A and B. Comparing our results to Wang et al. \cite{wang_prediction_2018} (Reference Model), we observe that the reference model experienced a larger MPS compared to all our models by up to 0.11, 0.10 and 0.03, respectively. It's important to note that modeling the ventricles as empty cavities in Model C results in a slightly smaller brain mass and moment of inertia compared to Models A and B. This reduction can lead to slightly higher brain acceleration for the same load applied to the skull in transient analysis. Consequently, we would anticipate potentially larger deformations and strains in Model C. This expectation aligns with our observations of higher maximum principal strain (Almansi) in Model C compared to Models A and B, as shown in Figure \ref{fig8}. Future studies could quantify the exact impact of this mass reduction on brain kinematics and resulting strains. 

\begin{figure*}
\centering 
\includegraphics[width=\textwidth]{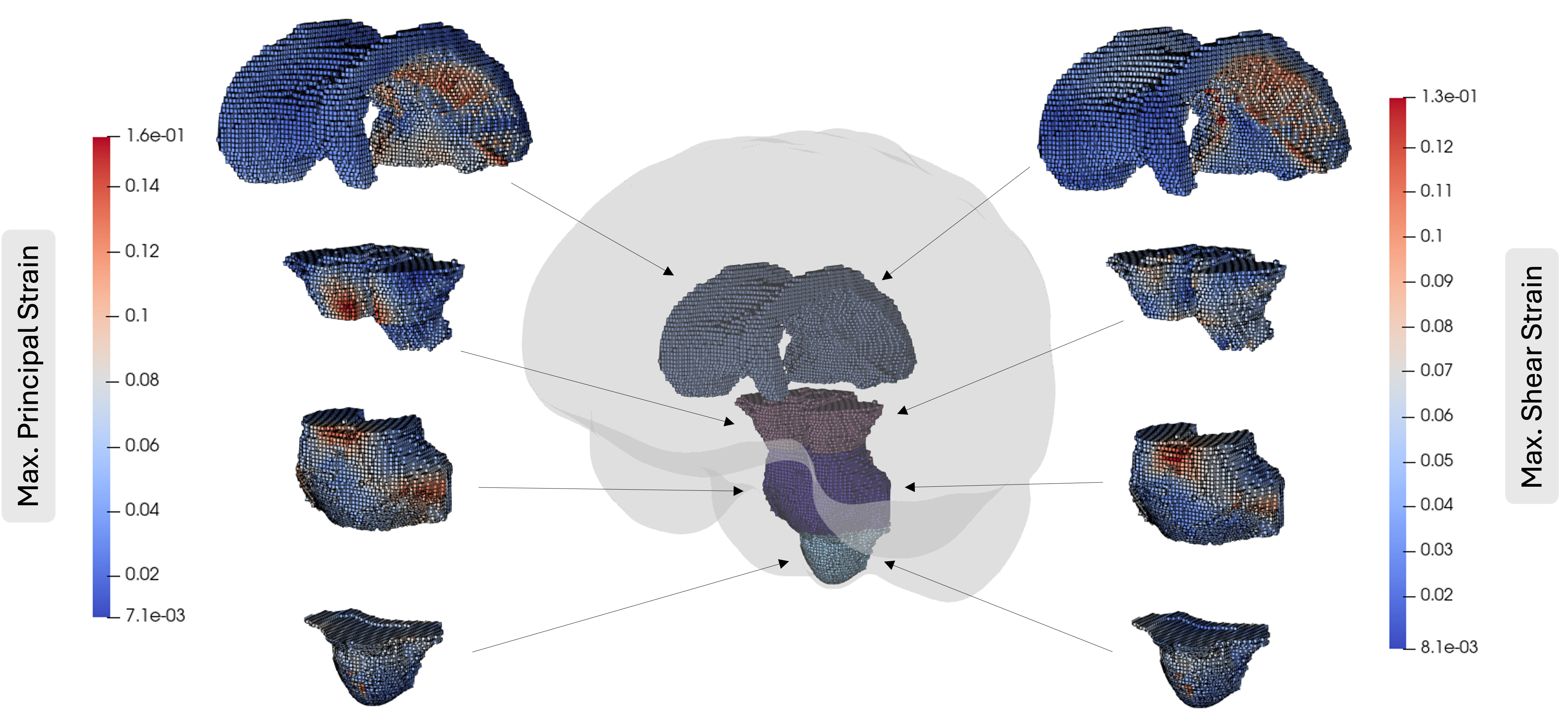}
\caption{The 3D model shows a snapshot of Model A and its important structures at the time when maximum strain value was observed in the case study simulation. The 3D model at the center is the meshed anatomical brain atlas used for the simulation, with important anatomical structures highlighted. These include the corpus callosum (grey), midbrain (pink), pons (dark blue) and medulla oblongata (light blue). The left side shows the maximum principal strain (Almansi) at the corresponding structures. The right side shows the maximum shear strain (Almansi) at the corresponding structures.} 
\label{fig9} 
\end{figure*}

In Figure \ref{fig9}, we analyse and visualise the computed maximum principal strain and maximum shear strain at important localised anatomical regions of the brain. This type of presentation of result is difficult to achieve without subject area knowledge and hence not typically used. However, we show that this is very easy to achieve when using the meshed anatomical atlas as the mesh's elements and nodes are labelled with the anatomical structures in the brain atlas. We extracted results for the corpus callosum, midbrain, pons and medulla oblongata. Statistics such as maximum, minimum and mean values of maximum principal and shear strain are reported in Tables \ref{tab3} and \ref{tab4}. Maximum values of MPS were found in the midbrain and medulla oblongata of 0.155, whereas maximumum values of shear strain was found in the corpus callosum of 0.170.

\begin{table*}[h!]
\caption{Statistics of maximum principal strain (Almansi) for important structures at the time when maximum strain value was observed.}
\begin{tabular*}{\textwidth}{@{\extracolsep\fill}llll}
    \toprule
    Part    &   Maximum &   Mean &   Minimum  \\
    \midrule
    Corpus callosum   &   0.136  &   0.0411    &   0.00487    \\
    Midbrain &   0.155  &   0.0377    &   0.00802  \\
    Pons    &   0.149  &   0.0549    &   0.0215   \\
    Medulla oblongata  &   0.155  &   0.0523    &   0.00462    \\
    \bottomrule
\end{tabular*}
\label{tab3}
\end{table*}

\begin{table*}[h!]
\caption{Statistics of maximum shear strain (Almansi) for important structures at the time when maximum strain value was observed.}
\begin{tabular*}{\textwidth}{@{\extracolsep\fill}llll}
    \toprule
    Part    &   Maximum &   Mean &   Minimum  \\
    \midrule
    Corpus callosum   &   0.170  &   0.048    &   0.00850    \\
    Midbrain &   0.110  &   0.0465    &   0.00367  \\
    Pons    &   0.132  &   0.0468    &   0.0134   \\
    Medulla oblongata  &   0.134  &   0.0480    &   0.00599    \\
    \bottomrule
\end{tabular*}
\label{tab4}
\end{table*}

\clearpage
\section{Case study II: Bioelectric field propagation}
Electroencephalography (EEG) has a wide variety of applications, with one of its most important uses being in the diagnosis and classification of epilepsy. Temporal lobe epilepsy (TLE), a common form of focal epilepsy, relies heavily on EEG for presurgical evaluation and diagnosis \cite{javidan_electroencephalography_2012, salami_seizure_2020}. EEG plays a crucial role in TLE treatment by providing information to predict responses to antiseizure drugs and to identify the seizure onset zone (SOZ) for surgical planning and resection. To locate the SOZ, a common method is to use the Finite Element Method (FEM) to solve the EEG forward problem. While there are many software options available for generating FE models and solving the EEG forward problem, they have limitations that our meshed anatomical atlas can address. Most existing software generates FE models using tetrahedral elements, which, although easily automated, lack the favorable numerical properties of regular hexahedra \cite{hughes_finite_2000}. Our meshed anatomical atlas incorporates geometry-adapted hexahedral elements, offering improved accuracy. Furthermore, models generated by current software \cite{fischl_freesurfer_2012, henschel_fastsurfer_2020, ashburner_unified_2005,gaser_cat_2024, riviere_brainvisa_2009} often have limited anatomical structures compared to OAP's SPL/NAC brain atlas, which restricts the detail when constructing the FE model and limits the scope for context-based analysis. We overcome this limitation by using OAP's SPL/NAC brain atlas, incorporating its comprehensive labels to create our mesh. Additionally, our mesh is compatible with various commercial \cite{ansys_ls-dyna_2021, smith_abaqusstandard_2009} and open-source software \cite{schroeder_visualization_2006, anderson_mfem_2021, fedorov_3d_2012, pieper_na-mic_2006}.

\subsection{Model description}
We present a workflow similar to our previous case study, constructing a bioelectrical model of the brain to solve the EEG forward problem using the meshed anatomical atlas. Our approach builds upon recent work \cite{zwick_patient-specific_2022, huynh_open_2024} that solves the EEG forward problem using the FE method. The EEG forward problem involves predicting the electric potential within the brain given a predefined source. In our model, we define the dipole source in the medial temporal region of the brain, specifically the left middle temporal gyrus, as epileptic seizure onset sources are typically modeled as current dipoles \cite{hallez_review_2007}. This simulation demonstrates the process of constructing a bioelectric model to solve the forward problem, a task that is particularly challenging due to the brain's numerous inhomogeneities and anisotropies \cite{barkley_meg_2003}. We implement the model and simulation using the open-source FE software MFEM \cite{anderson_mfem_2021}, leveraging its capabilities to handle our complex anatomical mesh and solve the bioelectric field equations efficiently.

\subsection{Modelling approaches}
\subsubsection{Governing equations of the brain bioelectric activity}
The physics of EEG can be approximated by Poisson's equation, which is the quasi-static approximation of Maxwell's equations. For spatial domain $\Omega \in \mathbb{R}^3$ with boundary $\partial \Omega = \overline{\Gamma_D \cup \Gamma_N}$ and outward unit normal $\boldsymbol{n}$, Poisson's equation for the EEG forward problem can be written as follows:
\begin{equation}
\label{eq3}
-\Delta \cdot (C(\nabla u)) = f \; \text{in} \; \Omega,
\end{equation}

\begin{equation}
\label{eq4}
\boldsymbol{n} \cdot (C(\nabla u)) = g \; \text{on} \; \Gamma_N,
\end{equation} 

\begin{equation}
\label{eq5}
u = h \; \text{on} \; \Gamma_D
\end{equation} 

where $u$ is the unknown scalar potential and $C$ is the (symmetric positive semi-definite) conductivity tensor. The low conductivity of air outside the scalp ($C = 0$ for all $x \notin \overline{\Omega}$) implies that a zero-flux Neumann boundary condition $g = 0$ can be applied on the surface $\Gamma_N$. Dirchlet boundary conditions $h$ on surface $\Gamma_D$ are typically applied by setting the potential to zero at a node corresponding to a reference electrode.

\subsubsection{Conductivity assignment}
We consider five tissue types for conductivity assignment: skull, scalp, white matter (WM), gray matter (GM) and cerebrospinal fluid (CSF). The isotropic conductivities were assigned as in Table \ref{table:conductivity}.

\begin{table*}[h!]
\caption{Conductive parts used in EEG forward model}
\begin{tabular*}{\textwidth}{@{\extracolsep\fill}lll}
    \toprule
    Part    &   Conductivity (S/m) & References  \\
    \midrule
    Scalp   &   0.33    &   Geddes and Baker (1967) \cite{geddes_specific_1967}; Stok (1987) \cite{stok_influence_1987}    \\
    Skull &   0.012 & Hallez et al. (2007) \cite{hallez_review_2007}; Gutierrez et al. (2004) \cite{gutierrez_estimating_2004}  \\
    Cerebrospinal fluid (CSF)    &   1.79   &   Baumann et al. (1997) \cite{baumann_electrical_1997}   \\
    Gray matter  &   0.33   & Geddes and Baker (1967) \cite{geddes_specific_1967}; Stok (1987) \cite{stok_influence_1987}    \\
    White matter    &   0.33    & Geddes and Baker (1967) \cite{geddes_specific_1967}; Stok (1987) \cite{stok_influence_1987}    \\
    \bottomrule
\end{tabular*}
\label{table:conductivity}
\end{table*}

\begin{figure*}
\centering 
\includegraphics[width=\textwidth]{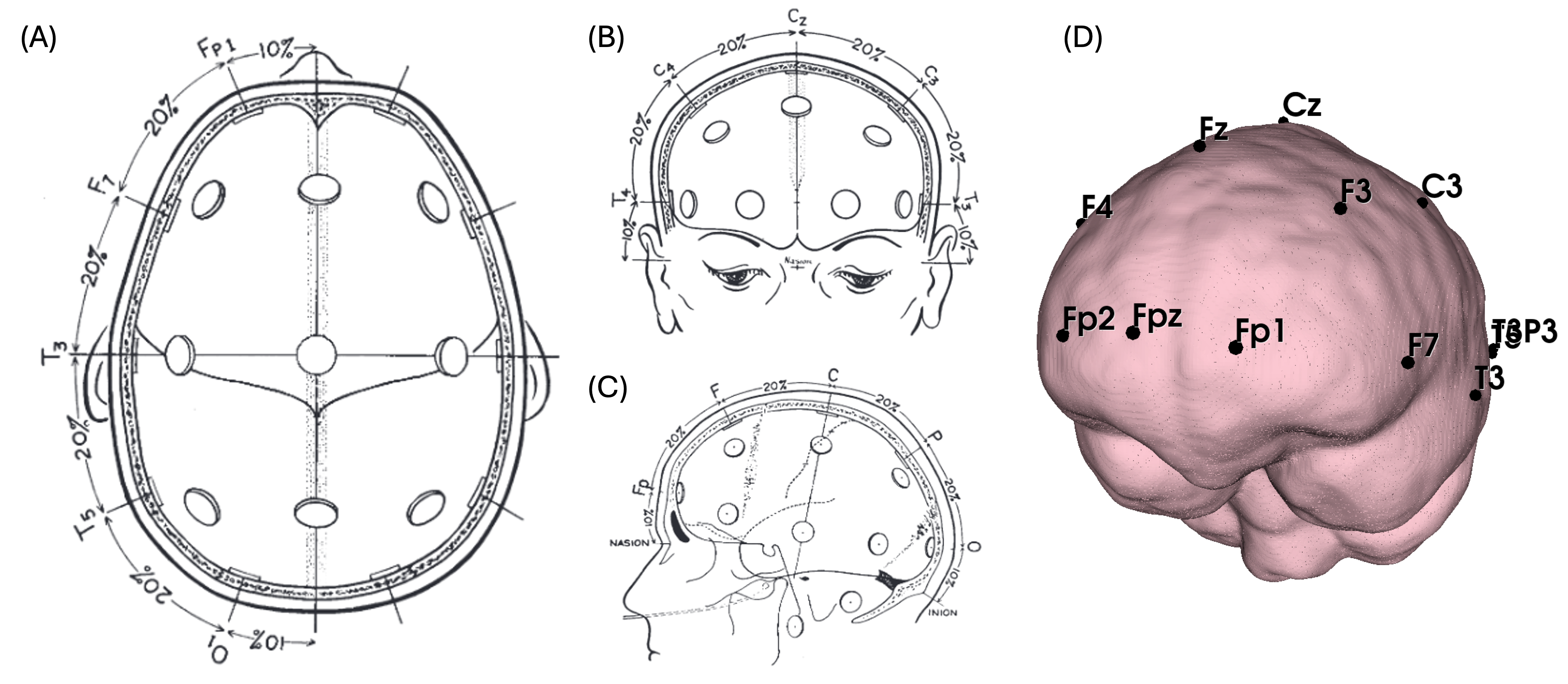}
\caption{The ten-twenty electrode system used to evaluate the electric potential generated by a current dipole. (A) Superior view (B) frontal view and (C) lateral view of skull showing the 10-20 electrode system (modified images from Klem et al. (D) 10-20 electrode system used for the case study.} 
\label{fig10} 
\end{figure*}

\subsection{Results of Bioelectric Field Propagation Case Study}

We present results similarly shown in our preliminary work \cite{huynh_open_2024}. This includes computed potentials overlaid onto both the labelled FE mesh and the volumetric MRI images with the SPL/NAC brain atlas label map. We projected electrodes onto the scalp surface following the international standard 10-20 electrode system \cite{klem_ten-twenty_1999} using open-source 3D Slicer extension, GeodesicSlicer \cite{briend_geodesicslicer_2020}, and evaluated the potentials at these locations.

\begin{figure*}
\centering 
\includegraphics[width=\textwidth]{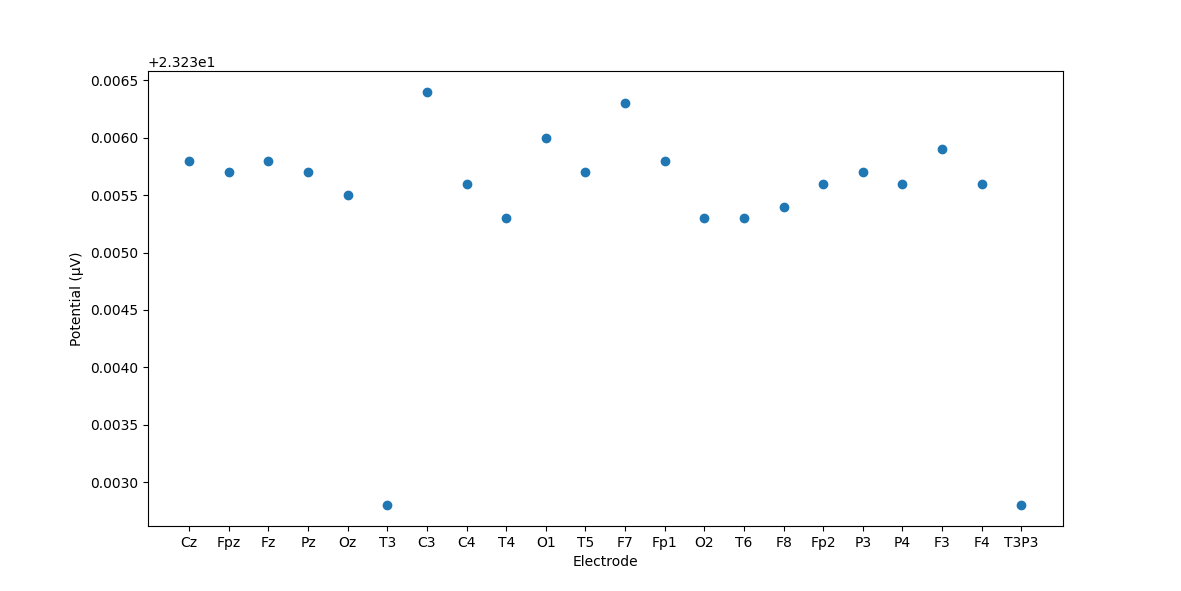}
\caption{Electrical potential in the brain generated by a current dipole located at the left middle temporal gyrus and evaluated at electrodes using the 10-20 electrode system.} 
\label{fig11} 
\end{figure*}

Figure \ref{fig11} illustrates the electric potential generated by a current dipole located in the left middle temporal gyrus, evaluated at electrodes using the 10-20 electrode system. The results show a notable decrease in potential at electrodes T3 and T3P3, corresponding to the simulated dipole location. This demonstration highlights the importance of accurately defining the positions of sensors and the source space for solving the forward problem, which ultimately yields the "leadfield matrix" mapping each dipole to every sensor \cite{medani_brainstorm-duneuro_2023}.

\begin{figure*}
\centering 
\includegraphics[width=\textwidth]{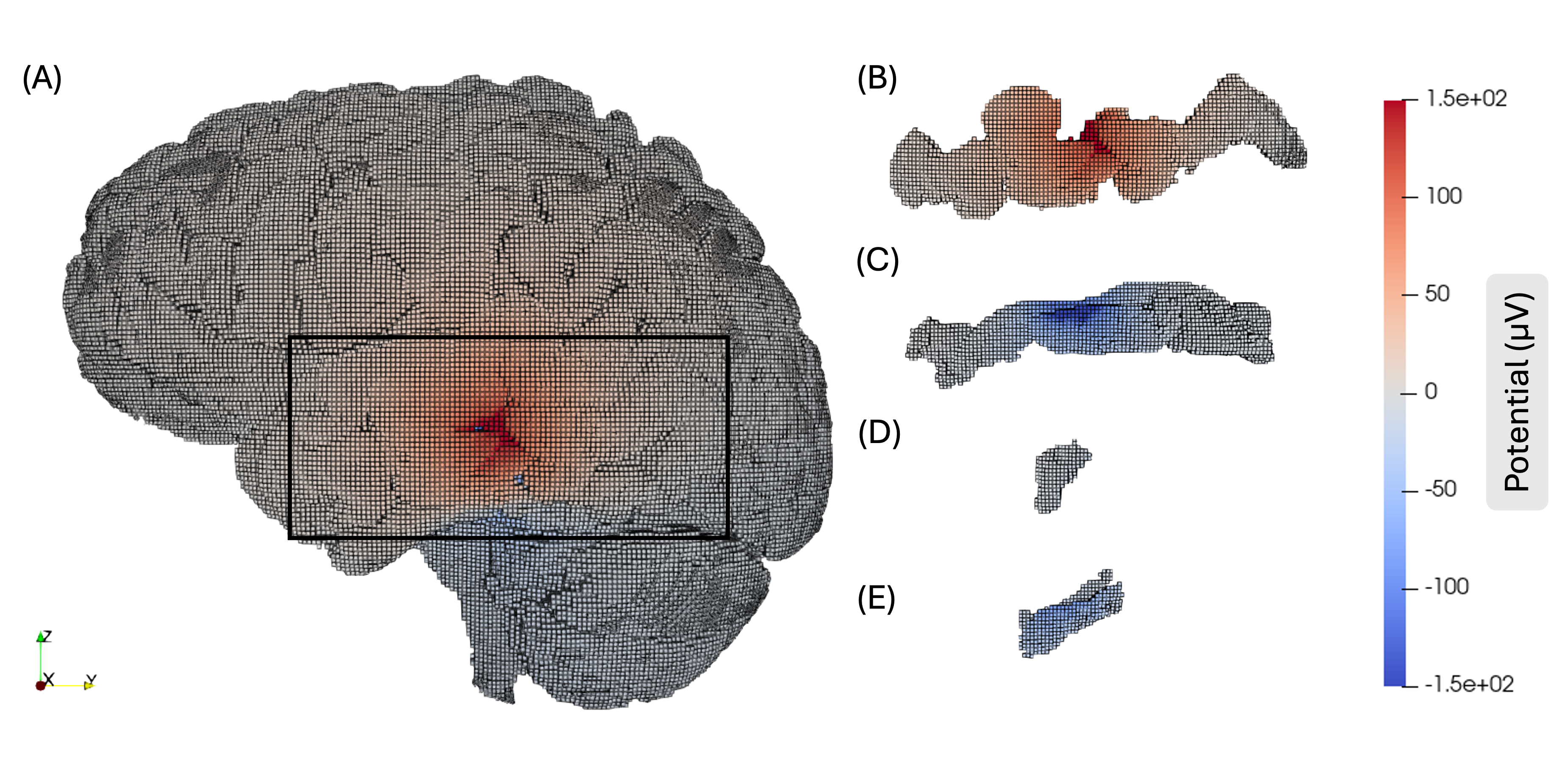}
\caption{Electric potential in the brain generated by a current dipole located at the left middle temporal gyrus overlaid onto the finite element (FE) mesh. (A) 3D side view of of the meshed anatomical brain atlas, (B) left middle temporal gyrus, (C) left amygdala (D) left parahippocampal gyrus (E) left fusiform gyrus.} 
\label{fig12} 
\end{figure*}

Figure \ref{fig12} displays the electric potential distribution within the brain, visualized on the FE mesh of the brain parenchyma with important anatomical structures isolated. Our labeled mesh allows for easy querying of specific anatomical structures corresponding to the OAP's SPL/NAC brain atlas, a process that would typically require extensive expertise in brain anatomy and effort to extract computed values at corresponding nodes and/or elements. We present the potential within key structures of the temporal lobe, including the left middle temporal gyrus, left amygdala, parahippocampal gyrus, and left fusiform gyrus. Table 6 shows the statistics of the electrical potential at these locations.

Furthermore, we demonstrate the capability to overlay computed potentials onto MRI scans and the OAP's SPL/NAC brain atlas label map. This feature may prove particularly useful for medical professionals in clinical settings, enabling them to correlate simulated electrical activity with anatomical structures more easily.

We demonstrate the versatility and potential of our meshed anatomical brain atlas in bioelectric field propagation modeling. The goal of our results is to show possible ways that researchers can benefit from using a FE mesh created from a comprehensive anatomical atlas. By combining detailed anatomical information with finite element modeling, we offer a powerful tool for more accurate and anatomically informed simulations of brain electrical activity.

\begin{table*}[h!]
\caption{Statistics of electric potential (\textmu V) at the left middle temporal gyrus, left amygdala, left parahippocampal gyrus and left fusiform gyrus.}
\begin{tabular*}{\textwidth}{@{\extracolsep\fill}llll}
    \toprule
    Part    &   Maximum (\textmu V) &   Mean (\textmu V) &   Minimum (\textmu V)  \\
    \midrule
    Left middle temporal gyrus   &   24368.6  &   43.96    &   -25749.3    \\
    Left amygdala &   0  &   -6.98    &   -16.01  \\
    Left parahippocampal gyrus    &   0  &   -25.89    &   -93.39   \\
    Left fusiform gyrus  &   1.21  &   -20.06    &   -212.93    \\
    \bottomrule
\end{tabular*}
\end{table*}

\begin{figure*}
\centering 
\includegraphics[width=\textwidth]{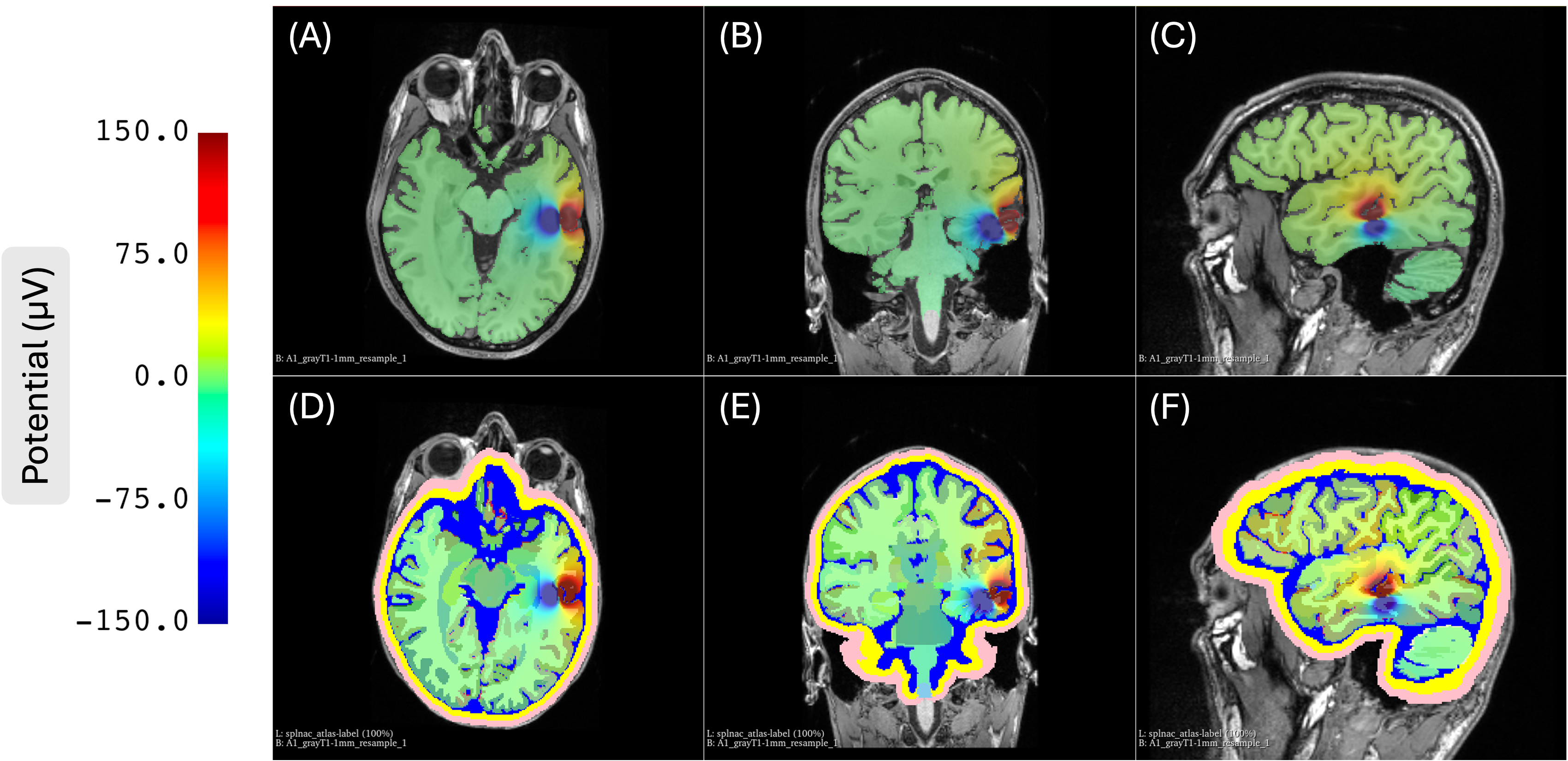}
\caption{Electric potential in the brain generated by a current dipole located at the left middle temporal gyrus overlaid onto the volumetric images in the axial, coronal and sagittal planes. Top row: Overlaid onto T-1 weighted MRI scan. Bottom row: Overlaid onto the OAP's SPL/NAC brain atlas label maps.}
\label{fig13} 
\end{figure*}

\section{Discussion}
\textbf{Overview.}
We have introduced a hexahedral mesh of an open-source digital anatomical atlas for construction of computational human brain models. Our work demonstrates the use of the overlay-grid (or mesh-first) method to generate a conforming hexahedral mesh, with smooth elements along important interfaces. Mesh quality for this method exhibit high quality for majority of the elements using metrics such as scaled Jacobian, aspect ratio, and skewness. We show that this method of generating hexahedral mesh can be easily modified to create varying levels of element count, depending on accuracy and computational requirements. We developed tools to edit the Open Anatomy Project's SPL/NAC anatomical atlases so that they can be adapted for finite element modeling. The elements of the conforming hexahedral mesh were then labeled using the OAP's SPL/NAC brain atlas label map and the Material label map (modified version) for constructing the FE models. We present two case studies: modeling biomechanics and bioelectric field propagation of the human brain. We show that efficient and accurate model construction was possible by referencing nodes or elements labeled with the anatomical structures for material property assignment, boundary conditions and surface contact definitions. Additionally, the integration of the anatomical structures with the mesh enables effective communication, visualisation and contextual analysis by relating anatomy with computed results. Our meshed anatomical brain atlas enables a seamless workflow for model creation, communication, analysis, and visualization in a relevant disciplinary context. 

\textbf{Significance of case study I and its results.}
In our first case study, we constructed a biomechanical model of the brain for impact injury. We build upon our previous research, which investigated modeling choices and their effects on brain responses under violent impact using a biomechanical model. A key improvement in our model is the inclusion of the ventricular system, which is often omitted in head models due to the time-consuming and difficult segmentation process. We constructed three different head models with varying properties of the ventricular system. We presented results typically reported in the field of injury mechanics of the brain, including quantile plots of the maximum principal and shear strains (Almansi), and the maximum principal strain (MPS). Additionally, we reported strain values at important anatomical locations, such as the corpus callosum, midbrain, pons and medulla oblongata, which are crucial for understanding diffuse axonal injury and its impact on cognitive function. In previous studies using biomechanical models to assess brain injury, global measures were often used. This made it difficult to assess and link mechanical behaviour and engineering outputs to brain behavior, cognitive impairment, or pathology. Our case study demonstrates the potential to enable more comprehensive studies in this area, moving beyond global measures to more specific, anatomically relevant analyses.

\textbf{Significance of case study II and its results.}
Our second case study focuses on constructing a model for simulating bioelectric field propagation by solving the EEG forward problem using FEM. Accurately solving the EEG forward problem is crucial for precise EEG source localization and analysis, which are essential in research and clinical applications for understanding the relationships between behavior and neural activity, as well as treating neurological disorders. We constructed the model with reference to our previous work \cite{zwick_patient-specific_2022}. To demonstrate the workflow and applicability of the meshed anatomical brain atlas in different disciplines and applications, we applied our modeling approach to represent a case of focal epilepsy. We showed that allocating a current dipole can be done accurately and intuitively by referencing the anatomical structures that the nodes and elements of the mesh are labeled with. We constructed a five-compartment mesh by modifying the SPL/NAC brain atlas labelmap using our SlicerAtlasEditor software extension, distinguishing between the skull, scalp, CSF, gray matter, and white matter. This distinction is highly recommended when modeling the head in EEG and MEG \cite{vorwerk_guideline_2014, cho_influence_2015}. We computed the electrical potential generated by a current dipole located in the left middle temporal gyrus of the brain and evaluated it at electrodes using the 10-20 electrode system, an internationally recognized standard for EEG electrode placement \cite{klem_ten-twenty_1999}. Our results show a large relative difference in potential at T3 and T3P3 electrodes, corresponding to the location of the simulated dipole. Additionally, by overlaying the computed electrical potential on both the labeled FE mesh and volumetric images (including the MRI and brain atlas label maps), users can query different anatomical structures and present statistics of computed results with respect to these structures. This approach offers researchers a better process to analyze and understand neural processes.

\section{Limitations and Future Work.}

While we have demonstrated the workflow for constructing FE head and brain models using the meshed anatomical atlas for two different research fields, there are several limitations in our current study that warrant addressing in future work.

\textbf{Subject-Specific and Patient-Specific Modeling.}
A primary limitation of this study is the inability to create subject/patient-specific FE head and brain models. This significantly constrains its applicability to subject and/or patient studies in neuroscience and clinical applications. Currently, many methods exists which can be implemented into workflows to create patient-specific geometry for FE models. One option is to use skull stripping. Skull stripping, also known as brain extraction, is a process which removes non-brain tissue signal from magnetic resonance imaging (MRI) data, leaving the brain. Skull stripping tools are widely explored and available in popular open source packages such as FreeSurfer \cite{fischl_freesurfer_2012}, however, the output is only sufficient to create the outer surface mesh of the brain. Additional steps are required to create a volumetric mesh suitable for the FEM. It also does not include any internal brain structures, requiring additional expertise and effort to segment. Another method that is used extensively by the computational neuroimaging community is constructing and using probabilistic brain atlases to automatically assign neuroanatomical labels to magnetic resonance images (MRI) of the brain \cite{fischl_automatically_2004}. This method is useful for constructing FE models as it automatically labels internal structures of the brain for property assignments. Additionally, the method works for unseen MRI images of the brain, allowing for construction of FE models for subject/patient-specific applications. A disadvantage of using this for our workflow is that the probabilistic atlases available do not cover as wide of a range of anatomical structures as the SPL/NAC brain atlas. The SPL/NAC brain atlas is also in continuous development by the community and medical experts, with new structures being added to the atlas. For the SPL/NAC brain atlas to be a probabilistic atlas, requires professional segmentation on a set of images for training. Additional structures added to the SPL/NAC atlas at a later date will require the training set to be updated with the new segments. As such, it is infeasible for such an atlas to be created and used for our workflow. As such, we think the best approach is to morph or register the existing meshed anatomical atlas to the individual's brain anatomy.

\textbf{Improving Representation of Head and Brain Model.}
Although the SPL/NAC brain atlas provides us with a comprehensive set of anatomical labels, it can still be improved by incorporating additional structures typically used in literature for modelling the brain. For example, research in the area of electrical modelling and simulation of the brain use highly detailed representation of the skull and scalp in their models \cite{huang_realistic_2019, nielsen_evaluating_2023}. Advanced segmentation processes are applied on individual's MRI, with recent research methods accurately segmenting the entire head for constructing head models \cite{puonti_accurate_2020, huang_realistic_2019, makarov_simnibs_2019}. Although this provides an accurate geometric representation of the head structures, the complex geometry makes it difficult for hexahedral meshing and can take several hours to complete \cite{nielsen_evaluating_2023}. In our workflow, we used an offset of the skullstrip surface to create a simplified skull and scalp. This simplified geometry helps the overlay-grid method produce high-quality hexahedral elements at the interfaces. There is a balance between lower quality elements using accurate geometries or high-quality elements using simplified geometries. As such, we think the best approach is to update the underlying SPL/NAC brain atlas label with both the accurate and simplified representations of additional head structures. In addition, structures that are often included in the FE model are the falx, tentorium and structures between the skull and the brain. These are currently not included in the atlas. Reasons may be due to the falx and tentorium having a thickness often less than the smallest voxel size in MRIs, making it difficult to detect and segment. Finding a way to include these structures would make our FE models more accurate. Lastly, it was found in our recent paper \cite{arzemanzadeh_towards_2024} that bridging veins can have effects on the brain-skull interface contact definition, restricting motion \cite{famaey_structural_2015}. Adding these would also improve our models.

\textbf{Element Count and Model Accuracy.}
While we have shown that our meshed anatomical atlas can be reduced in the number of elements, we did not demonstrate the difference in computed results between the model with one-to-one correspondence between voxels and elements and a reduced element count model. Conducing a sensitivity analyis to quantify the impact of element reduction on model accuracy and computational efficiency across various applications will be useful for applicability to real-world applications. 

\textbf{Software Compatibility.}
One of the aims for this project is to create a mesh that can be used in multidisciplinary studies. This means that the data format should be suitable for the corresponding software used in these areas. We have demonstrated compatibility with ANSYS LS-DYNA and open-source software and libraries such as 3D Slicer, VTK, and MFEM. To increase the utility of our proposed FE mesh, compatibility with other software packages should be considered. Currently, there exists open-source software and libraries for mesh conversion, such as meshio \cite{schlomer_meshio_2024}. Some are incorporated in FEA software such as LS-Dyna (LS-PrePost) and ABAQUS. We think the best approach is to create or contribute to an open-source conversion software which can seamlessly convert between mesh data formats. This will be beneficial to researchers to develop knowledge about the brain in multi-disciplines.
\section{Conclusions}
In this study, we have presented a finite element hexahedral mesh of the Open Anatomy Project's SPL/NAC brain atlas for multi-disciplinary scientific computation of the human brain. Our approach addresses the challenges posed by the complex structure and composition of the human brain, which typically requires detailed geometric and anatomical information for accurate modeling. By meshing the brain atlas directly, we bypass the need for manual segmentation of subject or patient MRI data, a process that is often time-consuming and error-prone. We have demonstrated a workflow for creating and editing these meshes to meet specific user requirements, offering a versatile foundation for various brain modeling applications.

Although our mesh does not conform to an individual's brain geometry, its primary strength lies in its broad applicability across multiple disciplines. We illustrated this versatility through two case studies: a biomechanical model predicting brain deformation and traumatic injury risk due to violent impact, and a bioelectric model solving the forward problem for source localization. These examples show the ease of constructing complex brain models using established approaches from literature, while also demonstrating the mesh's compatibility with software from different disciplines.

A key feature of our meshed anatomical atlas is the inclusion of labeled anatomical structures, which enables more insightful analysis and visualization of results. This labeling allows researchers to interpret findings in direct relation to brain anatomy, offering a level of contextual understanding that was challenging to achieve with previous unlabeled meshes. By providing this anatomical context, our work facilitates more meaningful analysis and interpretation of results across various neuroscience applications.

Our meshed anatomical brain atlas serves as a bridge between anatomical complexity and model construction, offering the community a readily available tool for diverse computational applications. It enables a seamless workflow for model creation, communication, analysis, and visualization in relevant disciplinary contexts, thereby advancing the field of scientific computing of the human brain. As we continue to refine this approach, future work will focus on developing methods for creating  subject-specific and patient-specific brain atlases, creating more accurate models by developing and adding structures to better represent the human head and brain, and expanding software compatibility. These advancements will further enhance the utility and impact of our meshed anatomical brain atlas in both research and clinical settings, contributing to a more comprehensive understanding of brain function and pathology. The meshed anatomical brain atlas is available on Github (https://github.com/andy9t7/MeshedAnatomicalAtlas).
\section*{Acknowledgments}

This research was carried out while the first author A.T. Huynh was in receipt of an “Australian Government Research Training Program Scholarship at The University of Western Australia”. Authors A. Wittek and K. Miller acknowledge funding from The University of Western Australia’s Research Income Growth Grant scheme for travel to Harvard University. This research was supported partially by the Australian Government through the Australian Research Council ARC Discovery Projects funding scheme (Project DP230100949). The authors acknowledge the support by Dr. Fang Wang (College of Automotive and Mechanical Engineering, Changsha University of Science and Technology) for providing state-of-the-art information on brain injury criteria.

\bibliographystyle{plainnat}

\nolinenumbers

\newpage
\listoffixmes

\end{document}